\documentclass[12pt]{article}
\usepackage{epsfig}
\usepackage{graphicx}
\usepackage{amssymb}
\usepackage{amsmath}
\usepackage{amsfonts}
\usepackage{mathrsfs}
\usepackage[dvips]{color}

\newcommand{\R}{\mathbb{R}}
\newcommand{\C}{\mathbb{C}}

\newcommand{\fz}{\mathfrak{z}}
\newcommand{\fA}{\mathfrak{A}}

\newcommand{\fD}{\mathfrak{D}}

\newcommand{\fK}{\mathfrak{K}}

\newcommand{\cH}{\mathcal{H}}

\newcommand{\be}{\begin{equation}}
\newcommand{\ee}{\end{equation}}
\newcommand{\bea}{\begin{eqnarray}}
\newcommand{\eea}{\end{eqnarray}}

\newcommand{\ed}{\end{document}}

\newcommand{\pbr}{\prec}
\newcommand{\pkt}{\succ}

\newcommand{\bi}{\begin{itemize}}
\newcommand{\ei}{\end{itemize}}

\newcommand{\bce}{\begin{center}}
\newcommand{\ece}{\end{center}}

\newcommand{\One}{{\stackrel{\leftrightarrow}{1}}}
\newcommand{\Ep}{{\stackrel{\leftrightarrow}{\mbox{\large{$\varepsilon$}}}}}
\newcommand{\MU}{{\stackrel{\leftrightarrow}{\mbox{\large{$\mu$}}}}}
\newcommand{\Mu}{{\stackrel{\;\leftrightarrow_\prime}{\mbox{\large{$\mu$}}}}}
\newcommand{\tEp}{{\stackrel{\leftrightarrow}{\mbox{\large{$\tilde\varepsilon$}}}}}
\newcommand{\tMU}{{\stackrel{\leftrightarrow}{\mbox{\large{$\tilde\mu$}}}}}

\newcommand{\sE}{\mathscr{E}}
\newcommand{\sB}{\mathscr{B}}

\begin{document}

\title{Pseudo-Hermiticity and Electromagnetic Wave \\ Propagation in
Dispersive Media}

\author{Ali~Mostafazadeh\thanks{E-mail address:
amostafazadeh@ku.edu.tr }
\\
Department of Mathematics, Ko\c{c} University, \\ 34450 Sar{\i}yer,
Istanbul, Turkey}

\date{ }
\maketitle

\begin{abstract}

Pseudo-Hermitian operators appear in the solution of Maxwell's
equations for stationary non-dispersive media with arbitrary
(space-dependent) permittivity and permeability tensors. We offer an
extension of the results in this direction to certain stationary
dispersive media. In particular, we use the WKB approximation to
derive an explicit expression for the planar time-harmonic solutions
of Maxwell's equations in an inhomogeneous dispersive medium and
study the combined affect of inhomogeneity and dispersion.

\vspace{2mm}

\vspace{2mm}

\end{abstract}
\vspace{5mm}

\section{Introduction}

A linear operator $A$ that acts in a Hilbert space is called
pseudo-Hermitian, if one can find an invertible Hermitian operator
$\eta$ satisfying $A^\dagger=\eta\,H\eta^{-1}$, \cite{p1}. This
notion of pseudo-Hermiticity arises naturally in the study of
non-Hermitian Hamiltonian operators, such as
$H=-\frac{d^2}{dx^2}+ix^3$, that admit a real spectrum
\cite{jpa-2006a}. It has also interesting applications in many
different areas \cite{review}. Perhaps one of the most remarkable of
these is the appearance of pseudo-Hermitian operators in classical
electrodynamics \cite{epl-2008}.

Consider the propagation of electromagnetic waves in a source-free
medium with arbitrary (possibly space-dependent) permittivity and
permeability tensors $\Ep$ and $\MU$. Then, the Maxwell equations
read
    \bea
    &&\vec\nabla\cdot\vec D=0,~~~~~~~~\vec\nabla\cdot\vec B=0,
    \label{max-1}\\
    &&\dot{\vec B}+\fD\vec E=0,~~~~~
    \dot{\vec D}-\fD \vec H=0,
    \label{max-2}
    \eea
where $\vec E$ and $\vec B$ are the electric and magnetic fields,
$\vec D=\Ep\vec E$, $\vec H=\MU^{-1}\vec B$, an over-dot means a
time-derivative, and $\fD$ is the curl operator, e.g., $\fD \vec E
:=\vec\nabla\times\vec E$.

For the cases that the medium is stationary, i.e., $\Ep$ and $\MU$
do not depend on time, we can reduce Eqs.~(\ref{max-2}) to
    \bea
    &&
    \vec B(\vec r,t)=\vec B_0(\vec r)-\int_0^t\fD\,\vec E(\vec
    x,\tau)\,d\tau,
    \label{B=}\\
    &&\ddot{\vec E}+\Omega^2\vec E=0,
    \label{wave-eq}
    \eea
where $\vec B_0(\vec r):=\vec B(\vec r,0)$ and
    \be
    \Omega^2:=\Ep^{-1}\fD\,\MU^{-1}\fD.
    \label{Omega}
    \ee
In view of (\ref{wave-eq}), the initial-value problem for Maxwell's
equations admits the formal solution \cite{epl-2008}:
    \be
    \vec E(\vec r,t)=\cos(\Omega t)\vec E_0(\vec r)+
    \Omega^{-1}\sin(\Omega t)\dot{\vec E}_0(\vec r),
    \label{E=}
    \ee
where $\vec E_0(\vec r):=\vec E(\vec r,0)$, and $\dot{\vec E}_0(\vec
x):=\dot{\vec E}(\vec r,0)=\Ep^{-1}\fD\,\MU^{-1}\vec B_0(\vec r)$.

We can view $\fD$, $\Ep$, $\MU$, $\Ep^{-1}$ and $\MU^{-1}$ as linear
operators acting in the vector space of complex-valued vector fields
$\vec F:\R^3\to\C^3$. Endowing this space with the $L^2$-inner
product, $\pbr \vec F,\vec G\pkt:=\int_{\R^3}\vec F(\vec r)^*\cdot
\vec G(\vec r)\,dx^3$, we find a Hilbert space $\cH$ in which the
curl operator $\fD$ acts as a Hermitian operator, i.e., for all
$\vec F,\vec G\in\cH$ belonging to the domain of $\fD$, we have
$\pbr \vec F,\fD\vec G\pkt=\pbr\fD\vec F,\vec G\pkt$. For a lossless
(and gainless) medium, the permittivity and permeability tensors
(and their inverses) also define Hermitian operators acting in
$\cH$. This in turn implies
$\Omega^{2\dagger}=\Ep\,\Omega^2\Ep^{-1}$, i.e., as an operator
acting in $\cH$, $\Omega^2$ is a pseudo-Hermitian operator.

In fact, because for a lossless medium $\Ep(\vec r)$ is a
positive-definite matrix for all $\vec r$, $\Ep$ is a
positive-definite operator, and $\Omega^2$ belongs to a special
class of pseudo-Hermitian operators, called quasi-Hermitian
\cite{quasi}. A basic property of quasi-Hermitian operators is that
they are related to Hermitian operators via similarity
transformations \cite{p2-p3}. For example, we can relate $\Omega^2$
to a Hermitian operator $h$ according to
    \be
    \Omega^2\longrightarrow h:=
    \Ep^{\frac{1}{2}}\,\Omega^2\,\Ep^{-\frac{1}{2}}=\Ep^{-\frac{1}{2}}
    \fD\:\MU^{-1}\fD\:\Ep^{-\frac{1}{2}}.
    \label{h=}
    \ee
This observation has been used in \cite{epl-2008} to devise a method
of solving the initial-value problem for Maxwell's equations. This
method applies to arbitrary stationary inhomogeneous and anisotropic
media, but it ignores the effects of dispersion. The purpose of the
present article is to extend the approach of \cite{epl-2008} to
stationary dispersive media.

\section{Time-Harmonic Solutions and Dispersion}

In standard textbook discussions of the propagation of the
electromagnetic waves in dispersive media, one usually begins by
assuming a harmonic time-dependence for the electromagnetic waves,
    \be
    \vec E(\vec r,t)= e^{-i\omega t}
    \vec \sE(\vec r) ,
    ~~~~
    \vec B(\vec r,t)= e^{-i\omega t}
    \vec \sB(\vec r) ,
    \label{htd}
    \ee
and considers the possibility that the permittivity and permeability
tensors $\Ep$ and $\MU$ depend also on the frequency $\omega$. In
Eqs.~(\ref{htd}), $\vec \sE$ and $\vec \sB$ are time-independent
complex-valued vector fields that in view of (\ref{max-1}) are
subject to the constraints
    \bea
    &&\vec\nabla\cdot(\Ep\,\vec\sE)=0,
    \label{div}\\
    &&\vec\nabla\cdot\vec\sB=0.
    \eea
Because according to (\ref{B=}) we can express the magnetic field in
terms of the electric fields, we will confine our attention to the
dynamics of the latter.

First, we consider the case of non-dispersive material.

Inserting the first of Eqs.~(\ref{htd}) in (\ref{wave-eq}) yields
    \be
    \Omega^2 \vec \sE  =\omega^2 \vec \sE.
    \label{eg-va}
    \ee
This reveals the interesting fact that the solutions of Maxwell's
equation having harmonic time-dependence are actually obtained from
the eigenfunctions of the pseudo-Hermitian operator $\Omega^2$.

In general, the eigenvalues of $\Omega^2$ are highly degenerate, and
the imposition of (\ref{div}) does not lift this degeneracy
completely. In the case that the wave propagates in vacuum, we have
$\Ep=\varepsilon_0\One$, $\MU=\mu_0\One$, and $\Omega^2=c^2\fD^2$.
Therefore, the eigenvalue equation (\ref{eg-va}) subject to the
condition~(\ref{div}) coincides with that of the Laplacian subject
to the same condition,
    \be
    -\nabla^2\vec\sE^\varnothing =
    k^2 \vec\sE^\varnothing ,
    ~~~~~~~~\vec\nabla\cdot\vec\sE^\varnothing =0.
    \label{vac}
    \ee
Here $k:=\omega/c$ and the superscript $\varnothing$ refers to the
vacuum solution. As is well-known, Eqs.~(\ref{vac}) admit the
following complete set of plane-wave solutions
    \be
    \vec\sE^\varnothing_{\vec k,\hat e_{\hat k}^\pm}(\vec r)
    :=\frac{e^{i\vec k\cdot\vec r}}{(2\pi)^{3/2}}\,\hat e^\pm_{\hat
    k},\label{vac-sol}
    \ee
where $\hat k$ and $\hat e^+_{\hat k}$ are any pair of orthogonal
unit vectors, $\hat e^-_{\hat k}:=\hat k\times\hat e^+_{\hat k}$,
and $\vec k:=k\hat k$. The unit vectors $\hat k$ and $\hat e^+_{\hat
k}$ play the role of the degeneracy labels for the eigenvectors.
Substituting (\ref{vac-sol}) for $\vec\sE(\vec r)$ in (\ref{htd}),
we find the following solutions of Maxwell's equations having
harmonic time-dependence
    \be
    \vec E^\varnothing_{\vec k,\hat e_{\hat k}^\pm}(\vec r,t)
    :=\frac{e^{i(\vec k\cdot\vec r-\omega t)}}{(2\pi)^{3/2}}\,
    \hat e^\pm_{\hat k}.
    \label{vac-sol-t}
    \ee
The general solution of Maxwell's equations in vacuum is a
superposition of these plane wave solutions,
    \be
    \vec E^\varnothing(\vec
    x,t)=\int_{\R^3}d^3\vec k\int_0^{2\pi}d\varphi_{\vec k}\left[
    c_+(\vec k,\varphi_{\vec k})
    \vec E^\varnothing_{\vec k,\hat e_{\hat k}^+(\varphi_{\vec k})}(\vec
    x,t)+c_-(\vec k,\varphi_{\vec k})
    \vec E^\varnothing_{\vec k,\hat e_{\hat k}^-(\varphi_{\vec k})}(\vec
    x,t)\right].
    \label{sup-vac}
    \ee
Here for each $\vec k\neq\vec 0$ we fix a polar coordinate system
$(\rho_{\vec k},\varphi_{\vec k})$ in the plane normal to $\vec k$,
$\hat e_{\hat k}^+(\varphi)$ is the unit vector specified by the
polar angle $\varphi_{\vec k}$ in this plane,
 $\hat e_{\hat k}^-(\varphi_{\vec k})=\hat k\times\hat e_{\hat
 k}^+(\varphi_{\vec k})$, and $c_\pm$ are complex-valued mode functions.

Now, consider the more general case that the medium is locally
inhomogeneous and/or anisotropic. In this case, we will suppose that
the degeneracy structure of the eigenvectors of $\Omega^2$ is the
same as in the case of vacuum. Therefore, we denote these
eigenvectors by $\vec\sE_{\vec k,\hat e_{\hat k}^\pm}$, and express
the corresponding solutions of Maxwell's equations with harmonic
time-dependence (\ref{htd}) as
    \be
    \vec E_{\vec k,\hat e_{\hat k}^\pm}(\vec r,t)=e^{-i\omega t}
    \vec\sE_{\vec k,\hat e_{\hat k}^\pm}(\vec r).
    \label{ths-2}
    \ee
It is usually customary to take the general solution of Maxwell's
equations to be a superposition of these time-harmonic solutions.

For the case that the medium is lossless, $\Omega^2$ is a
quasi-Hermitian operator with a complete set of eigenfunctions
$\vec\sE_{\vec k,\hat e_{\hat k}^\pm}$. This ensures the existence
of an eigenfunction expansion for $\Omega^2$ and provides the
mathematical justification for the claim that the most general
solution $\vec E$ of Maxwell's equations is a superpositions of
$\vec E_{\vec k,\hat e_{\hat k}^\pm}$,
    \be
    \vec E(\vec r,t)=\int_{\R^3}d^3\vec k\int_0^{2\pi}d\varphi_{\vec k}\left[
    c_+(\vec k,\varphi_{\vec k})
    \vec E_{\vec k,\hat e_{\hat k}^+(\varphi_{\vec k})}(\vec
    x,t)+c_-(\vec k,\varphi_{\vec k})
    \vec E_{\vec k,\hat e_{\hat k}^-(\varphi_{\vec k})}(\vec
    x,t)\right].
    \label{sup-vac-gen}
    \ee

Next, consider the case of a stationary dispersive medium. Then, one
usually accounts for the $\omega$-dependence of $\Ep$ and $\MU$ in
determining the time-Harmonic solutions $\vec E_{\vec k,\hat e_{\hat
k}^\pm}$ and  considers the superposition of these solutions to
describe waves having more general initial conditions, e.g.,
localized wave packets. This requires solving (\ref{eg-va}). Note,
however, that the operator $\Omega^2$ depends on $\omega$.
Therefore, this equation can no longer be viewed as a standard
eigenvalue equation. Yet one can still perform the transformation
(\ref{h=}) to determine an $\omega$-dependent operator $h$ and
express the solutions of Eq.~(\ref{eg-va}) in terms of those of
    \be
    h\,\vec\psi =\omega^2\vec\psi,
    \label{eg-va-h-1}
    \ee
according to
    \be
    \vec\sE =\Ep^{-\frac{1}{2}}\vec\psi.
    \label{eg-ve}
    \ee

In the absence of dispersion, $h$ is a genuine Hermitian operator
with a complete orthonormal set of eigenvectors.
Ref.~\cite{epl-2008} uses this observation to evaluate the
right-hand side of the formal solution of Maxwell's equation given
by (\ref{E=}). Here we explore the consequences of allowing $\Ep$,
$\MU$, and $h$ to depend on $\omega$. Because a direct extension of
the results of \cite{epl-2008} to general dispersive media is beyond
our knowledge, we will confine our attention to separable cases
where
    \be
    \Ep(\vec r,\omega)=\varepsilon_1(\omega)\Ep_2(\vec r),~~~~~~~~~~~~
    \MU(\vec r,\omega)=\mu_1(\omega)\MU_2(\vec r),
    \label{sep}
    \ee
$\varepsilon_1,\mu_1$ are positive real-valued functions of
$\omega$, and $\Ep_2,\MU_2$ are positive $3\times 3$ matrix-valued
functions of $\vec r$.\footnote{In terms of the transformed
permeability and permittivity tensors, $\tEp(t,t',\vec r,\vec
r^{\:\prime})$ and $\tMU(t,t',\vec r,\vec r^{\:\prime})$, that
appear in the constitutive relations \cite{AG}: $\vec D(\vec
r,t)=\int_{-\infty}^t dt'\int_{\R^3}d^3\vec r^{\:\prime}\:
\tEp(t,t',\vec r,\vec r^{\:\prime})\vec E(\vec
r^{\:\prime},t')$,\linebreak $\vec B(\vec r,t)=\int_{-\infty}^t
dt'\int_{\R^3}d^3\vec r^{\:\prime}\:
    \tMU(t,t',\vec r,\vec r^{\:\prime})\vec H(\vec r^{\:\prime},t'),$
The separability condition (\ref{sep}) takes the form
$\tEp(t,t',\vec r,\vec r^{\:\prime})=\tEp_1(t-t')\tEp_1(\vec r)
    \delta(\vec r^{\:\prime}-\vec r)$, $\tMU(t,t',\vec r,\vec r^{\:\prime})=\tMU_1(t-t')\tMU_1(\vec r)
    \delta(\vec r^{\:\prime}-\vec r)$.}
The separability condition~(\ref{sep}) is not quite realistic, but
as we will see in the following, it provides us with a concrete
analytically treatable class of toy models. Furthermore, we can use
it to establish the completeness of the eigenfunctions of $h$.

For the cases for which (\ref{sep}) holds, the $\omega$-dependence
of $\Omega^2$ and $h$ factors, and we have
    \be
    \Omega^2=n_1(\omega)^{-2}\Omega^2_2,~~~~~~h=n_1(\omega)^{-2}h_2,
    \label{factor}
    \ee
where
    \be
    n_1(\omega):=\sqrt{\varepsilon_1(\omega)\mu_1(\omega)},~~~~~
    \Omega^2_2:=\Ep_2^{-1}\fD\,\MU_2^{-1}\fD,~~~~~
    h_2:=
    \Ep_2^{\frac{1}{2}}\,\Omega_2^2\,\Ep_2^{-\frac{1}{2}}=\Ep_2^{-\frac{1}{2}}
    \fD\:\MU_2^{-1}\fD\:\Ep_2^{-\frac{1}{2}}.
    \label{x-ind}
    \ee
As seen from (\ref{x-ind}), $h_2$ is a Hermitian
($\omega$-independent) operator acting in $\cH$, and the
eigenfunctions $\vec\psi$ of $h$ with eigenvalue $\omega^2$ coincide
with the eigenfunctions $\vec\psi_2$ of $h_2$ with eigenvalue
$\omega^2n_1(\omega)^2$. Because $h_2$ is Hermitian, the latter form
a complete orthonormal basis of $\cH$. This in turn implies that the
eigenfunctions of $h$ will also form a complete orthonormal basis of
$\cH$ provided that the function
    \be
    f(\omega):=\omega\, n_1(\omega)
    \label{f=}
    \ee
is a monotonically increasing function of $\omega$ and we normalize
the eigenfunctions properly. This together with (\ref{eg-ve})
constitute a mathematical basis for expanding the general solution
of Maxwell's equations in terms of time-Harmonic solutions for
dispersive media satisfying (\ref{sep}).

\section{Planar Waves in an Inhomogeneous Dispersive Medium}

Consider an isotropic but inhomogeneous dispersive medium defined by
    \be
    \Ep=\varepsilon(z,\omega)\One,~~~~~~~~~~~~\Mu=\mu(z,\omega)\One,
    \label{ze1}
    \ee
where $\varepsilon$ and $\mu$ are positive real-valued functions.
Suppose that a linearly polarized planar wave propagates along the
$z$-axis in this medium in such a way that both the electric and
magnetic fields are independent of the $x$- and $y$-coordinates. The
initial data have the form
    \be
    \vec E_0(\vec r)=\vec E_0(z)={\cal E}_0(z)\hat i,~~~~~
    \vec B_0(\vec r)=\vec B_0(z)={\cal B}_0(z)\hat j,
    \label{ini-condi}
    \ee
where $\hat i$ and $\hat j$ are respectively the unit vectors along
the $x$- and $y$-axes.

Introducing $p:=-i\frac{d}{dz}$, we can express the operators
$\Omega^2$ and $h$ as
    \be
    \Omega^2=\varepsilon(z,\omega)^{-1}p\,\mu(z,\omega)^{-1}p,~~~~~
    h=\varepsilon(z,\omega)^{-1/2}p\,\mu(z,\omega)^{-1}p\;
    \varepsilon(z,\omega)^{-1/2}.
    \ee
Therefore, similarly to the non-dispersive case considered in
\cite{epl-2008}, Eq.~(\ref{eg-va-h-1}) reads as
    \be
    -\frac{1}{\sqrt{\varepsilon(z,\omega)}}\frac{d}{dz}\left[
    \frac{1}{\mu(z,\omega)}\frac{d}{dz}\left(\frac{\psi(z)}{
    \sqrt{\varepsilon(z,\omega)}}\right)\right]=\omega^2\psi(z).
    \label{eg-va-h-2}
    \ee
Here we have dropped the vector sign from $\vec\psi$, because it is
always parallel to the $x$-axis; $\vec\psi(z)=\psi(z)\hat i$.

Eq.~(\ref{eg-va-h-2}) is the time-independent Schr\"odinger equation
for a point particle moving along the $z$-axis and having a
position- and energy-dependent-mass. Following~\cite{epl-2008}, we
will employ WKB approximation to solve this equation. This yields
solutions of the form
    \be
    \tilde\psi_{\omega}(z):=
    \frac{\nu(\omega)\:e^{i\omega u(z,\omega)}}{\sqrt{v(z,\omega)}},
    \label{wkb-dis}
    \ee
where $\nu(\omega)$ are nonzero complex normalization constants, and
    \be
    u(z,\omega):=\int_0^z \frac{d\fz}{v(\fz,\omega)},~~~~~~
    v(z,\omega):=\frac{1}{\sqrt{\varepsilon(z,\omega)\mu(z,\omega)}} .
    \label{u-v=}
    \ee
Substituting (\ref{wkb-dis}) for $\vec\psi$ in (\ref{eg-ve}) and
using (\ref{ze1}), we find a set of solutions $\vec\sE_\omega$ of
Eq.~(\ref{eg-va}). Using these in (\ref{htd}) yields the
time-harmonic solutions for the system. In order to address the
completeness of these solutions, we should address the completeness
of (\ref{wkb-dis}). We are able to do this for the separable cases
where
    \be
    \varepsilon(z,\omega)=\varepsilon_1(\omega)\varepsilon_2(z),
    ~~~~~~~~~~
    \mu(z,\omega)=\mu_1(\omega)\mu_2(z).
    \ee
The result is the following WKB-approximate eigenfunctions of $h$.
    \be
    \psi_{\omega}(z):=
    \sqrt{\frac{f'(\omega)}{2\pi\,v_2(z)}}
    \;e^{if(\omega) u_2(z)},
    \label{wkb-dis-comp}
    \ee
where $f$ is the function given in (\ref{f=}), a prime stands for a
derivative, and
    \be
    v_2(z):=\frac{1}{
    \sqrt{\varepsilon_2(z)\mu_2(z)}},~~~~~~~~~~
    u_2(z):=\int_0^z \frac{d\fz}{v_2(\fz)}.
    \ee
Under the assumption that $f$ is a monotonically increasing
function, we can easily establish the following orthonormality and
completeness relations for $\psi_{\omega}$.
    \be
    \int_{-\infty}^\infty
    \psi_{\omega}(z)^*\psi_{\omega'}(z)\;dz=\delta(\omega-\omega'),~~~~~
    \int_{-\infty}^\infty
    \psi_{\omega}(z)\psi_\omega(z')^*\;d\omega=\delta(z-z').
    \label{orth-comp}
    \ee

Having obtained a complete set of eigenfunctions of $h$, we can
construct time-Harmonic WKB-approximate solutions of Maxwell's
equations. In view of (\ref{htd}) and (\ref{eg-ve}), these have the
form
    \be
    \vec E_\omega(z,t)=\frac{e^{-i\omega
    t}\psi_{\omega}(z)\,\hat i}{
    \sqrt{\varepsilon_1(\omega)\varepsilon_2(z)}}=
    \left(\frac{\mu_1(\omega)\mu_2(z)}{\varepsilon_1(\omega)
    \varepsilon_2(z)}\right)^{1/4}
    \sqrt{\frac{1}{2\pi}\left[1+\frac{\omega\,n_1'(\omega)}{n_1(\omega)}\right]}
    \;\exp\left\{i\omega[n_1(\omega) u_2(z)- t]\right\}\hat i.
    \label{htd-wkb}
    \ee
Note that the condition of the validity of the WKB approximation is
the same as in the non-dispersive case ($\varepsilon_1=\mu_1=1$)
discussed in \cite{epl-2008}, namely
    \be
    \frac{v_2^2}{2}\left|
    \frac{2v_2v_2''-{v_2'}^2}{2v_2^2}+\frac{2\mu_2\mu_2''-3{\mu_2'}^2}{2\mu_2^2}
    \right|\ll\omega^2,
    \label{condi}
    \ee
where we have suppressed the $z$-dependence of $v_2$ and $\mu_2$ for
simplicity.

In light of (\ref{condi}), the general WKB solution of Maxwell's
equations with initial conditions of the form (\ref{ini-condi})
reads
    \be
    \vec E(z,t)=\int_{-\infty}^\infty c(\omega)\vec E_\omega(z,t)\,
    d\omega,
    \label{gen-sol}
    \ee
where the modulus of the mode function $c(\omega)$ is supposed to be
negligibly small for all $\omega$ violating (\ref{condi}). We can
use (\ref{orth-comp}), (\ref{htd-wkb}), and (\ref{gen-sol}) to
express the mode function $c(\omega)$ in terms of the initial
electric field. This yields
    \be
    c(\omega)=\sqrt{\varepsilon_1(\omega)}
    \int_{-\infty}^\infty
    \sqrt{\varepsilon_2(z)}\;\psi_\omega(z)^*{\cal E}_0(z)~dz,
    \label{c=}
    \ee
where ${\cal E}_0(z):=\vec E(z,0)\cdot\hat i$. Therefore, the
condition of the validity of the WKB approximation restricts the
choice of the initial condition.

If the medium happens to be homogeneous, i.e., $\varepsilon_2$ and
$\mu_2$ are constants, the WKB approximation is exact, $v_2$ is
constant, $u_2$ takes the form $u_2(z)=z/v_2$, and the time-harmonic
solution (\ref{htd-wkb}) coincides with a plane wave,
$A\,e^{i(kz-\omega t)}$, where
    \bea
    A&:=&\left(\frac{\mu(\omega)}{\varepsilon(\omega)}\right)^{1/4}
    \sqrt{\frac{1}{2\pi}\left[1+\frac{\omega\,n_1'(\omega)}{n_1(\omega)}
    \right]},
    \label{amp}\\
    k&:=&\omega n_1(\omega)/v_2=\omega\sqrt{\varepsilon(\omega)\mu(\omega)}.
    \label{wn}
    \eea
Therefore, if we consider the time-evolution of an initial plane
wave:
    \be
    {\cal E}_0(z)=A~e^{ikz},
    \label{pw}
    \ee
we will not be able to observe the effect of dispersion. The
situation is the opposite in an inhomogeneous dispersive medium.
Because of the $z$-dependence of $v_2$ and $\varepsilon_2$,
inserting (\ref{pw}) in (\ref{c=}) gives a mode function $c(\omega)$
that is not proportional to a Dirac delta function. Consequently,
the shape of the field changes in time; \emph{in an inhomogeneous
medium an initial plane wave feels the effect of dispersion}.

For example, consider a nonmagnetic material ($\mu_1(\omega)=1$,
$\mu_2(z)=\mu_0$) with a Lorentzian inhomogeneity:
    \be
    \varepsilon_2(z)=\varepsilon_0
    \left(1+\frac{a}{1+z^2/\gamma^2}\right),~~~~~a,\gamma\in\R^+,
    \label{Lorentz}
    \ee
and an arbitrary dispersion relation given by
$\varepsilon_1(\omega)$. Then to the first order in the strength of
the inhomogeneity $a$, the mode function (\ref{c=}) corresponding to
the plane wave (\ref{pw}) takes the form
    \be
    c(\omega)=\fA(\omega)\left\{\delta(k-\fK(\omega))+
    \left[\frac{e^{-\gamma|k-\fK(\omega)|}(3k-\fK(\omega))}{8(k-\fK(\omega))}
    \right]\gamma a\right\}+{\cal O}(a^2),
    \label{c=1}
    \ee
where
    \be
    \fA(\omega):=A\sqrt{\left(\frac{2\pi
    \varepsilon_0\:\varepsilon_1(\omega)^{3/2}}{c}\right)
    \left(1+\frac{\omega\varepsilon'_1(\omega)}{2\varepsilon_1(\omega)}
    \right)},~~~~~~~
    \fK(\omega):=\frac{\omega\sqrt{\varepsilon_1(\omega)}}{c}.
    \ee
The term in the square bracket in (\ref{c=1}) reflects the
dispersion of the plane wave (\ref{pw}) due to the inhomogeneity. It
is interesting to see that this term is exponentially suppressed for
the wave numbers violating the dispersion relation $k=\fK(\omega)$.

\section{Conclusion} In this paper we have examined the role of
pseudo-Hermitian operators in the description of electromagnetic
waves propagating in a stationary, possibly inhomogeneous or
anisotropic dispersive medium.

In the absence of dispersion the properties of the pseudo-Hermitian
operator $\Omega^2$ associated with the Maxwell equations lead to a
powerful spectral method for solving the initial-value problem for
these equations \cite{epl-2008}. In particular, for an effectively
one-dimensional model, it yields an explicit expression for the
propagating wave provided that one can employ the WKB approximation.
When the medium is dispersive the same approach can not be pursued.
Nevertheless, the pseudo-Hermitian operator $\Omega^2$ still
generates the time-harmonic solutions as its eigenfunctions. This
raises the question of the completeness of these eigenfunctions. In
this article we addressed this question for the cases that the
frequency- and the space-dependence of the permittivity and
permeability tensors are separable. Although this condition was
rather unrealistic, it allowed for a concrete implementation of our
general method. In particular, for an effectively one-dimensional
(planar) system with this separability property, we were able to
construct a complete set of WKB-approximate eigenfunctions of
$\Omega^2$ and study the combined effect of dispersion and
inhomogeneity showing that unlike for a homogenous medium in an
inhomogeneous medium a plane wave also undergoes dispersion.

\section*{Acknowledgments} I would like to express my gratitude to
Emre Kahya and Giuseppe Scolarici for many helpful discussions. This
work has been supported by the Scientific and Technological Research
Council of Turkey (T\"UB\.{I}TAK) in the framework of the project
no: 108T009, and by the Turkish Academy of Sciences (T\"UBA).

\ed